\documentclass[aps,prl,twocolumn,showpacs,reprintnumbers,amsmath,amssymb,superscriptaddress,citeautoscript]{revtex4-1}
\usepackage{graphicx}
\usepackage{dcolumn}

\begin{document}

\title{Probing itinerant antiferromagnetism with $d$-wave Andreev reflection spectroscopy}%

\author{C. R. Granstrom}
\affiliation{Department of Physics, University of Toronto, Toronto, Ontario M5S 1A7 Canada}

\author{R.-X. Liang}
\affiliation{Department of Physics and Astronomy, University of British Columbia, Vancouver, British Columbia V6T 1Z1 Canada}
\affiliation{Canadian Institute for Advanced Research, Toronto, Ontario M5G 1Z8, Canada}

\author{Y. Li}
\affiliation{Department of Materials Science and Engineering, University of Toronto, Ontario M5S 3E4, Canada}

\author{P. Li}
\affiliation{Department of Materials Science and Engineering, University of Toronto, Ontario M5S 3E4, Canada}

\author{Z. - H. Lu}
\affiliation{Department of Materials Science and Engineering, University of Toronto, Ontario M5S 3E4, Canada}

\author{E. Svanidze}
\affiliation{Department of Physics and Astronomy, Rice University, Houston, Texas 77005, USA}

\author{E. Morosan}
\affiliation{Department of Physics and Astronomy, Rice University, Houston, Texas 77005, USA}

\author{J. Y.T. Wei}
\affiliation{Department of Physics, University of Toronto, Toronto, Ontario M5S 1A7 Canada}
\affiliation{Canadian Institute for Advanced Research, Toronto, Ontario M5G 1Z8, Canada}

\begin{abstract}
To study how Andreev reflection (AR) is affected by itinerant antiferromagnetism, we perform $d$-wave AR spectroscopy with superconducting YBa$_2$Cu$_3$O$_{7-\delta}$ on TiAu and on variously-oxidized Nb (NbO$_x$) samples. X-ray photoelectron spectroscopy is also used on the latter to measure their surface oxide composition. Below the N\'eel temperatures ($T_N$) of both TiAu and NbO$_x$, the conductance spectra show a dip-like structure instead of a zero-bias peak within the superconducting energy gap; for NbO$_x$, higher-oxidized samples show a stronger spectral dip at zero bias. These observations indicate that itinerant antiferromagnetic order suppresses the AR process.  Interestingly, the spectral dip persists above $T_N$ for both TiAu and NbO$_x$, implying that spin fluctuations can also suppress AR. Our results suggest that $d$-wave AR spectroscopy may be used to probe the degree of spin ordering in itinerant antiferromagnets.
\end{abstract}

\maketitle

There is general interest in the interplay of superconductivity and itinerant magnetism, both on a fundamental level and for technological applications \cite{Buzdin2005,Dai2012}. At a normal-metal/superconductor (N/S) interface, Andreev reflection (AR) is the process that converts electrons into Cooper pairs through retro-reflection of holes \cite{Andreev1964,Blonder1982}. There has been a considerable amount of theoretical and experimental work on AR in itinerant ferromagnet/superconductor interfaces \cite{Jong1995,Upadhyay1998,Soulen1998,Ji2001,parker2002,Zutic2004,Buzdin2005,Nadgorny2011,Turel2011}, where AR has been utilized to probe spin-polarization. In contrast, there has been very little work on AR for itinerant antiferromagnet/superconductor (IAFM/S) interfaces, aside from the theoretical work in Refs. \cite{Andersen2002,Andersen2005,Bobkova2005}. However, there are topics of fundamental and applied importance that motivate study of such interfaces. For example, the interplay between antiferromagnetism and superconductivity is thought to be important in the high critical temperature ($T_c$) superconducting cuprates \cite{Moriya1990,Monthoux1992,Monthoux1994}, and in the high-$T_c$ iron-pnictides there is thought to be coexistence of superconducting and spin-density wave (SDW) states \cite{Hirschfeld2011,Dai2012}. On the technological side, Josephson junctions involving IAFMs are predicted to exhibit unique properties \cite{Gorkov2001}, and it is desirable to find new probes to characterize IAFMs, e.g. to use for antiferromagnetic spintronics applications \cite{Gomonay2017,Baltz2018}.

There has been recent theoretical work for AR onto IAFMs, and in particular, AR with $d$-wave superconductors (dSCs) \cite{Andersen2002,Andersen2005,Bobkova2005}. In analogy with AR in N/S junctions, a N/IAFM junction is predicted to exhibit a spin-dependent Q-reflection, whereby the SDW gap $\Delta_{\mathrm{SDW}}$ in the IAFM plays the role of the superconducting gap. In Q-reflection, quasiparticles with energy $E<\Delta_{\mathrm{SDW}}$ and momentum $\mathbf{k}_F$ incident upon the IAFM undergo a spin-dependent retroreflection into states with momentum $\mathbf{k}_F+\mathbf{Q}$, where $\mathbf{Q}$ is the AFM wavevector \cite{Andersen2005,Bobkova2005}. If Q-reflection is combined with AR in an dSC/IAFM junction, a variety of low-energy interfacial bound states are predicted to form, showing up as peaks in the differential conductance spectrum.

$d$-wave AR spectroscopy using superconducting YBa$_{2}$Cu$_{3}$O$_{7-\delta}$ (YBCO) tips and films is potentially a powerful technique for probing IAFMs, as it is expected to give spin-sensitive information on electronic states at the sub-nanometer length scale over a wide temperature range ($\sim 0-90$ K) \cite{Turel2011,Granstrom2018}. The sensitivity to spin polarization of such $d$-wave AR measurements was demonstrated with nanoscale YBCO point contacts on both Au and CrO$_2$, the latter showing suppression of $d$-wave AR as expected for a half-metallic ferromagnet \cite{Turel2011}, and recently with non-contact tip-sample junctions onto another half-metallic ferromagnet La$_{2/3}$Ca$_{1/3}$MnO$_{3}$ (LCMO), which also showed suppression of $d$-wave AR \cite{Granstrom2018}. 

\begin{table*}[ht]
\centering
\renewcommand{\arraystretch}{1.2}
\begin{tabular}[c]{c|c|c|c|c}
    Compound                       & Oxidation state      & $\rho$ ($\Omega$ cm)  & $\Theta_{\mathrm{CW}}$ (K) & $T_N$ (K) \\
\hline 
Nb$_{2}$O$_{5}$      (NbO$_{2.5}$)   & +5 	      & $3\times10^{4}$           & --- & --- \\
Nb$_{25}$O$_{62}$    (NbO$_{2.48}$)  & +5 	      & $3\times10^{-1}$           & 0 & --- \\
Nb$_{47}$O$_{116}$   (NbO$_{2.468}$) & +5 	      & $1.6\times10^{-2}$          & 7 & ---         \\
Nb$_{22}$O$_{54}$    (NbO$_{2.455}$) & +5 	      & $1.5\times10^{-2}$         & 12 & --- \\
Nb$_{12}$O$_{29}$    (NbO$_{2.417}$) & +5 	      & $4\times10^{-3}$          & 24 & 12\\
NbO$_{2}$                            & +4 	      & $6.3\times10^{3}$ \cite{Janninck1966}         & --- & --- \\
NbO                                  & +2 	      & $2.1\times10^{-5}$ \cite{Hulm1972}        & --- & --- \\
\end{tabular}
\caption{\label{tab:NbOx} Nb oxides, their electrical resistivity $\rho$ at 300 K, Curie-Weiss temperatures $\Theta_{\mathrm{CW}}$, and N\'eel temperatures $T_N$. For non-stoichiometric compounds, the listed oxidation state is the predominant one. Of the Nb$_2$O$_{5-x}$ compounds, our AR data is likely predominantly sensitive to Nb$_{12}$O$_{29}$, since AR is only known to be sensitive to itinerant moments. Adapted from Ref. \cite{Cava1991}.}
\end{table*}

In this work, we perform $d$-wave AR spectroscopy with superconducting YBCO to probe itinerant antiferromagnetism in two systems. One system is TiAu, a recently-discovered IAFM with no magnetic constituents \cite{Svanidze2015}. The other system is the Nb oxides (NbO$_x$), which have tunable antiferromagnetic order  \cite{Cava1991} and potentially itinerant antiferromagnetism \cite{McQueen2007}, both arising from oxygen vacancies. X-ray photoelectron spectroscopy (XPS) is also used on the latter to measure their surface oxide composition. Below the N\'eel temperatures ($T_N$) of both TiAu and NbO$_x$, the conductance spectra show a dip-like structure instead of a zero-bias peak within the superconducting energy gap; for NbO$_x$, higher-oxidized samples show a stronger spectral dip at zero bias. These observations indicate that itinerant antiferromagnetic order suppresses the AR process.  Interestingly, the spectral dip persists above $T_N$ for both TiAu and NbO$_x$, implying that spin fluctuations can also suppress AR. Our results suggest that $d$-wave AR spectroscopy may be used to probe the degree of spin ordering in itinerant antiferromagnets.

TiAu antiferromagnetically orders below 36 K, and XPS data suggests that Ti is close to its non-magnetic $4+$ oxidation state, ruling out the presence of local moments \cite{Svanidze2015}. Muon spin-relaxation data indicates 100\% volume fraction of magnetic order at 0 K and strong spin-fluctuations, but the exact role and strength of the latter are not currently known. Neutron diffraction measurements indicate long-range antiferromagnetic order, with a small itinerant moment of 0.15 $\mu_B$ per Ti atom. The fact that TiAu has no magnetic constituents defies existing theories \cite{Svanidze2015}. 

The most stable of NbO$_x$, Nb$_{2}$O$_{5}$, is electrically insulating when pure. However, it tends to have \textit{local} oxygen vacancies \cite{Grundner1984,Halbritter1987}, creating local moments via Nb$^{4+}$ ions. Additionally, the constituent NbO$_{6}$ octahedra of Nb$_{2}$O$_{5}$ can accommodate extended oxygen vacancies via crystallographic shear \cite{VanLanduyt1974,Nico2016}, forming several ordered and non-stoichiometric Nb$_{2}$O$_{5-x}$ compounds (table \ref{tab:NbOx}). As $x$ increases in Nb$_{2}$O$_{5-x}$, electrical conductivity and antiferromagnetic coupling between local moments increase, until finally Nb$_{12}$O$_{29}$ antiferromagnetically orders with a N\'eel temperature of 12 K. Of the Nb$_{2}$O$_{5-x}$ compounds, only Nb$_{12}$O$_{29}$ is metallic, and it is generally believed that some of its Nb$^{4+}$ sites have localized magnetic electrons, while other sites have itinerant and non-magnetic electrons \cite{Cava1991,Waldron2004,Ohsawa2011}. However, Nb$_{12}$O$_{29}$ may also be an IAFM, as recent work suggests its antiferromagnetism comes from delocalized electrons \cite{McQueen2007}.

Proslier and Cao et al. recently studied the influence of Nb$_{2}$O$_{5}$ on the superconductivity of Nb using XPS, point-contact tunneling (PCT) with normal-metal tips, and magnetic susceptibility measurements on NbO$_x$ samples \cite{Proslier2008,Proslier2008a,Proslier2009,Proslier2011,Cao2014}. Their PCT spectra showed signs of antiferromagnetic exchange-scattering between tunneling electrons and local moments in oxygen-deficient Nb$_{2}$O$_{5}$, as described by the Anderson-Appelbaum theory \cite{Appelbaum1966,Anderson1966,Appelbaum1967}. Magnetic susceptibility vs. temperature measurements on these same samples showed Curie-Weiss-like temperature dependence with positive Curie temperatures, also indicating antiferromagnetic coupling \cite{Proslier2011,Cao2014}.  The success of these normal-metal tip PCT studies in detecting signatures of antiferromagnetism motivates us to extend them to using superconducting $d$-wave YBCO tips, where the high $T_c$ of 90 K allows spin ordering to be probed over a wide temperature range.

In the present work, $d$-wave AR spectroscopy measurements were made using point-contact and scanning tunneling spectroscopy geometries. These measurements were made using a home-built $^4$He dipper-probe scanning tunneling microscope (STM), designed to allow tip and sample loading within a glovebox filled with dry N$_2$ gas. Measurement electronics are described elsewhere \cite{Granstrom2018,Granstrom2016}. XPS measurements were performed with a PHI 5500 analytical chamber using monochromated Al K$_{\alpha}$ radiation of 1486.7 eV at a take-off angle of 75$^{\circ}$ (for details see Ref. \cite{Adinolfi2016}).

The YBCO single crystals used as tips were grown using the self-flux technique \cite{Liang2012}. The preparation of the crystals for use as a tip is described elsewhere \cite{Granstrom2018}. The Nb tips were cut from 0.25mm diameter wire and were passively oxidized by leaving in air for an hour. The Nb foils used as samples were 25 $\mu$m thick and of 99.8\% purity. The Nb films used as samples were 30-200 nm thick and were grown on natively-oxidized Si wafers by sputtering. To provide a control sample for AR measurements, 40 nm thick Ag films were grown on mica by sputtering. Natively-oxidized Nb samples were reduced either by sputtering with 0.5-3 keV Ar ions in a background pressure of 0.1 $\mu$Torr, or annealing between 880-930 $^{\circ}$C under a vacuum of 2-4 $\mu$Torr for 25-105 hours. Immediately following each reduction treatment, samples were kept in a dry N$_2$ environment before quickly being mounted either onto the STM inside the glovebox for AR spectroscopy measurements, or into a vacuum chamber for XPS measurements. The (110)-YBCO films were 100 nm thick and grown by pulsed laser deposition on (110)- (LaAlO$_{3}$)$_{0.3}$(Sr$_{2}$TaAlO$_{6}$)$_{0.7}$ (LSAT) substrates, and were similarly transported in an N$_2$ filled container before being loaded onto the STM.

The TiAu polycrystals were grown by arcmelting \cite{Svanidze2015}, and were cut into $\sim$ 5 x 5 x 2 mm$^3$ slabs for use as a sample, or $\sim$ 1 x 1 x 5 mm$^3$ bars for use as a tip. The slabs were polished to a mirror finish, cleaned with Ar ion sputtering, then quickly transferred to an N$_2$ filled glovebox, where they were loaded onto the STM. The bars were broken inside the glovebox to expose a fresh edge, before being mounted onto the STM. Since both TiAu and Nb have chemically reactive surfaces at room temperature in air, various tip/sample combinations were measured.

\begin{figure}
\includegraphics[width=0.49\textwidth]{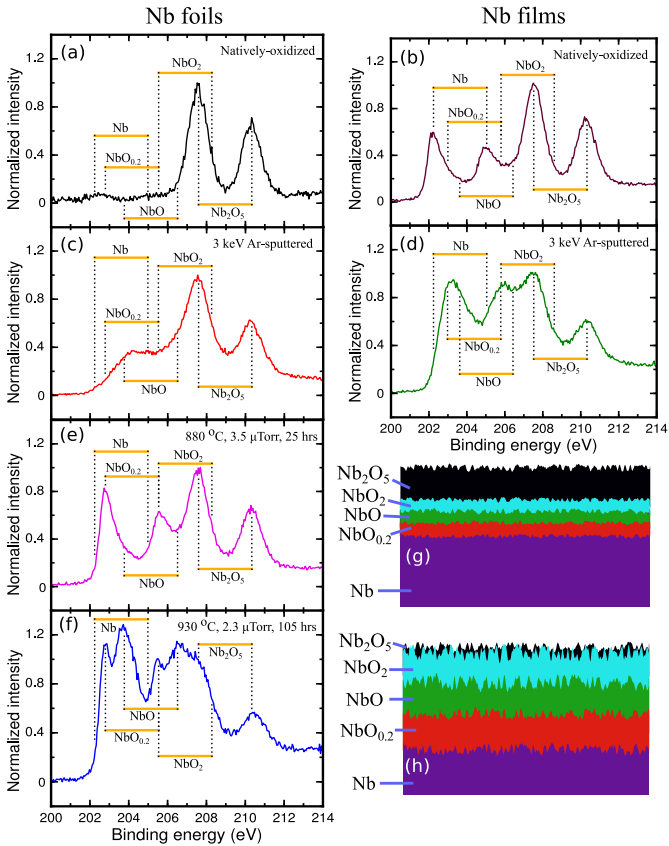}
\caption{\label{fig:XPS} XPS measurements of oxidized Nb samples. (a) and (b): The surface of natively-oxidized samples is predominantly Nb$_{2}$O$_{5}$, as schematically shown in (g). Layer roughness indicates non-uniformity in oxide depth composition \cite{Darlinski1987}. (c) and (d): Ar-ion sputtering thins the Nb$_{2}$O$_{5}$ surface layer by injecting oxygen into bulk Nb, thickening the lower oxidation states beneath the surface as shown in (h). (e) and (f): Vacuum annealing reduces the surface oxides similar to Ar-sputtering. Intensity is normalized to the 207.55 eV peak of Nb$_{2}$O$_{5}$. Colors of curves match corresponding samples in figure \ref{fig:YBCO_PCS_Nb_Ag}(a).
}
\end{figure}

\begin{figure}
\includegraphics[width=0.45\textwidth]{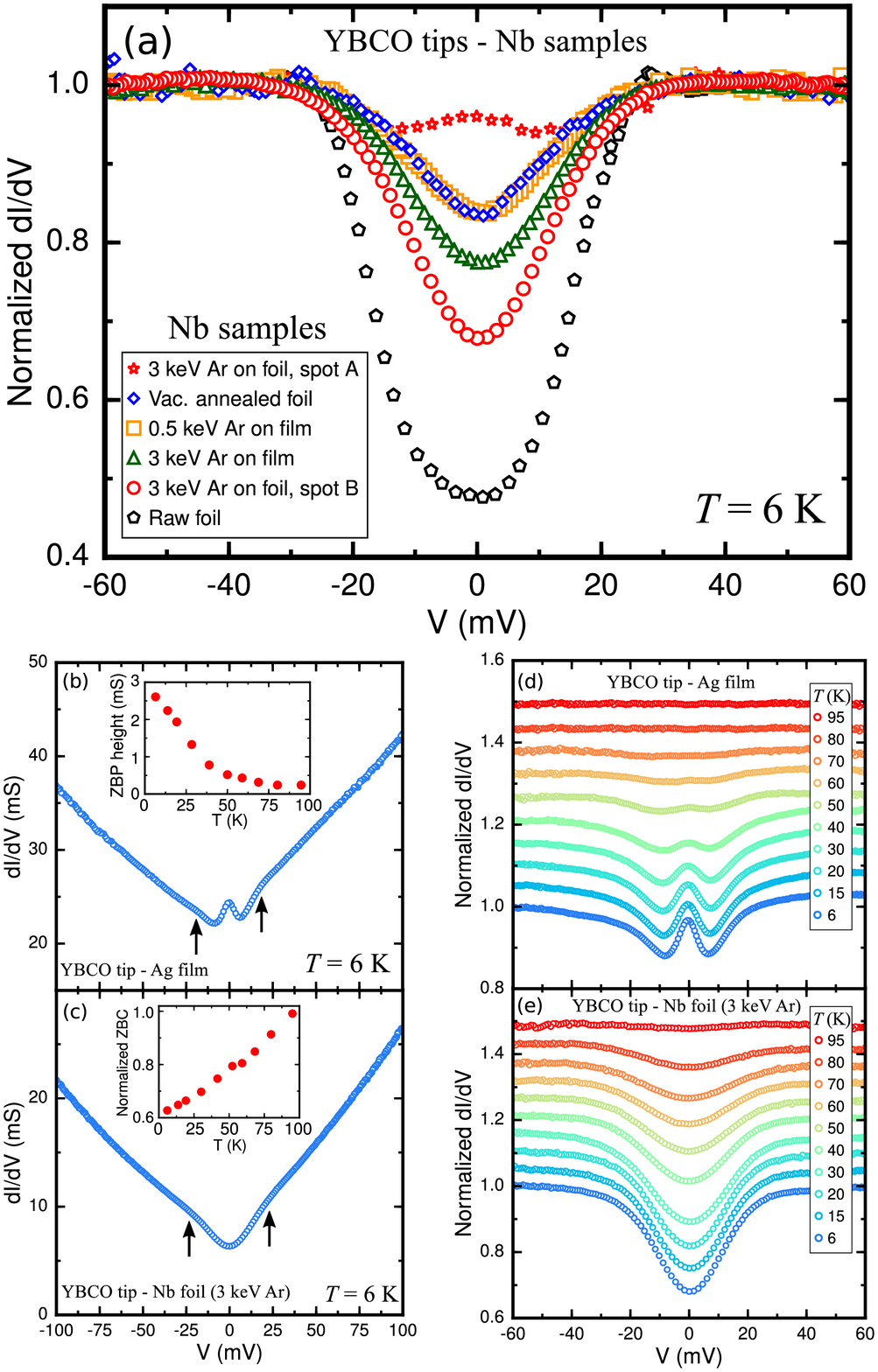}
\caption{\label{fig:YBCO_PCS_Nb_Ag} (a) Normalized point-contact spectra at 6 K between YBCO tips and NbO$_x$ samples reduced similarly to the NbO$_x$ samples in figure \ref{fig:XPS} (colors of data match figure \ref{fig:XPS}). The depth of the ZBC dip here and the Nb$_{2}$O$_{5}$ spectral weight in figure \ref{fig:XPS} are clearly correlated. (b) Raw data at 6 K on a Ag film and (c) Nb foil. Both spectra exhibit a similar conductance background and peaks near $\pm$ 20 mV, but the Ag data show a ZBC peak while the Nb data show a ZBC dip. (d) and (e): Temperature dependence of normalized data from panels (b) and (c), respectively. The ZBC features weaken at higher temperatures and disappear at YBCO's $T_{c}$ of 90 K. Insets of panels (b) and (c) show temperature evolution of ZBC for panels (d) and (e), respectively.
}
\end{figure}

We first demonstrate control over the surface oxide composition of NbO$_x$ with XPS, which is sensitive to the top $\sim$10 nm of our samples' surfaces \footnote{$\gtrsim 95$\% of photoelectrons come from within 3 inelastic mean free paths of the sample surface. Using the 1486.7 eV energy of Al K$_{\alpha}$ x-rays, the inelastic mean free path of photoelectrons in our samples ranges from 2.5 nm in Nb to 3.9 nm in Nb$_{2}$O$_{5}$, calculated using the expressions from Ref. \cite{Seah1979}.}. Figure \ref{fig:XPS} shows XPS measurements of the Nb 3d core level. Panels (a) and (b) show that the surface of both natively-oxidized foils and films is predominately Nb$_{2}$O$_{5}$, as expected. The film data show asymmetric Nb peaks, indicating the presence of lower Nb oxidation states beneath the surface Nb$_{2}$O$_{5}$, as shown schematically in panel (g) and reported by others \cite{Darlinski1987}. Note that it is difficult to distinguish stoichiometric Nb$_{2}$O$_{5}$ from non-stoichiometric Nb$_{2}$O$_{5-x}$ compounds with XPS \cite{Ohsawa2011}, since the predominant oxidation state is +5 in all compounds. Nevertheless, in practice numerous oxygen vacancies exist in Nb$_{2}$O$_{5}$ \cite{Grundner1984,Halbritter1987,Proslier2011}, so our large Nb$_{2}$O$_{5}$ peaks suggest that a combination of Nb$_{2}$O$_{5}$, non-stoichiometric Nb$_{2}$O$_{5-x}$ compounds, and locally oxygen-deficient Nb$_{2}$O$_{5}$ are present in our samples. 

The XPS spectra in figures \ref{fig:XPS}(c) and (d) show that sputtering with 3 keV Ar ions shifts spectral weight from Nb$_{2}$O$_{5}$ to lower Nb oxidation states, reducing the film more than the foil. Ar-sputtering oxidized Nb is believed to both preferentially remove oxygen from Nb$_{2}$O$_{5}$ and diffuse oxygen into bulk Nb \cite{Karulkar1981}, thickening the lower oxidation states beneath the surface as shown schematically in panel (h). Panels (e) and (f) show that vacuum annealing has a similar effect as Ar-sputtering: heating oxidized Nb to $\sim900$ $^{\circ}$C in low oxygen partial pressures diffuses oxygen atoms from Nb$_{2}$O$_{5}$ into bulk Nb, while temperatures above 1600 $^{\circ}$C are needed to completely evaporate the oxygen \cite{Strongin1972}. These XPS data clearly demonstrate that we can significantly thin the native Nb$_{2}$O$_{5}$ surface layer of Nb samples with Ar-sputtering and vacuum annealing.

Next, we show that the reduction treatments in figure \ref{fig:XPS}(c)-(f) enhance AR. Figure \ref{fig:YBCO_PCS_Nb_Ag}(a) shows point-contact spectra taken with YBCO tips on Nb samples oxidized similarly to those in figure \ref{fig:XPS}, where the data colors match between the two figures. The depth of the ZBC dip in figure \ref{fig:YBCO_PCS_Nb_Ag}(a) and the Nb$_{2}$O$_{5}$ spectral weight in figure \ref{fig:XPS} are clearly correlated. Reduced samples occasionally exhibit ZBC peaks (stars), suggesting spatial variation of the surface oxide composition in the reduced samples. The point-contact spectrum for the 0.5 keV Ar-sputtered film (squares) shows a higher ZBC than the 3 keV Ar-sputtered film (triangles), consistent with lower-energy Ar ions more effectively reducing Nb$_{2}$O$_{5}$ \cite{Karulkar1981}. Estimates of the tip-sample junction sizes using the Wexler formula \cite{Wexler1966} show that all junctions are ballistic \footnote{The resistance $R$ of a point-contact can be related to the contact's radius $a$ with the Wexler formula \cite{Wexler1966}:  $R=4\rho\ell/(3\pi a^{2})+\rho/2a$, where $\rho$ is electrical resistivity and $\ell$ is electronic mean free path. For our $R$ of 0.15-100 k$\Omega$, we use YBCO's $\ell$ of 10 nm and normal state $\rho$ of 50 $\mu\Omega$ cm \cite{Wei1998a} to estimate that $a\approx$0.2-5 nm, well below 10 nm.}, indicating that the ZBC features in figure \ref{fig:YBCO_PCS_Nb_Ag}(a) are not due to junction imperfections.

\begin{figure}
\includegraphics[width=0.4\textwidth]{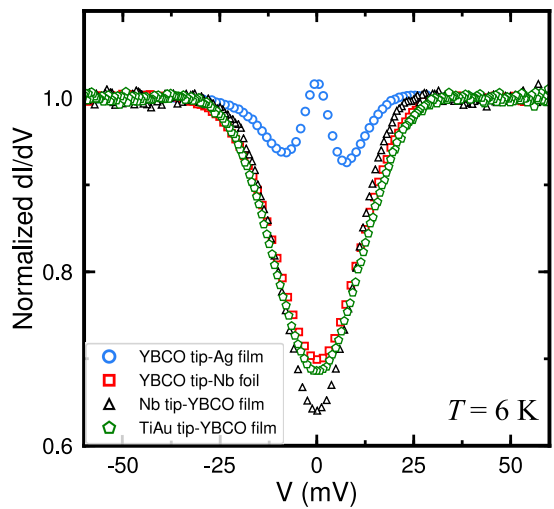}
\caption{\label{fig:YBCO_PCS_Nb_TiAu} Normalized point-contact spectra taken between YBCO and both NbO$_x$ and TiAu. For comparison, a spectrum for a YBCO tip onto an Ag film is also shown. Both the Nb and TiAu samples show a strong ZBC dip, in contrast to the ZBC peak in the YBCO-Ag spectrum.}
\end{figure}

To confirm that the ZBC dips in figure \ref{fig:YBCO_PCS_Nb_Ag}(a) are from AR suppression, we compare raw point-contact spectra taken on a Ag film and an Ar-sputtered Nb foil in figures \ref{fig:YBCO_PCS_Nb_Ag}(b) and (c), respectively. The YBCO-Ag data show a prominent zero-bias conductance (ZBC) peak, flanked by weak peaks near $\pm$ 20 mV. Similar ZBC peaks have been observed in a variety of point-contact studies on YBCO, primarily on crystal faces not normal to a principal crystal axis, and attributed to $d$-wave Andreev resonance \cite{Deutscher2005}. As for the side peaks, their energies are comparable to the superconducting gap maximum of YBCO \cite{Wei1998a, Sharoni2001, Ngai2007}. A quasi-linear conductance background is also present, as commonly observed for YBCO and other cuprates \cite{Geerk1988,Sun1994,Wei1998}. The YBCO-Nb foil data show similar features, except a ZBC dip is present instead of a ZBC peak. The temperature dependence of the ZBC features in these junctions is shown in their respective insets, which were extracted from the normalized data in panels (d) and (e). As temperature increases, the ZBC features weaken and finally disappear above YBCO's $T_{c}$ of 90 K, confirming that they are related to AR.

\begin{figure}
\includegraphics[width=0.45\textwidth]{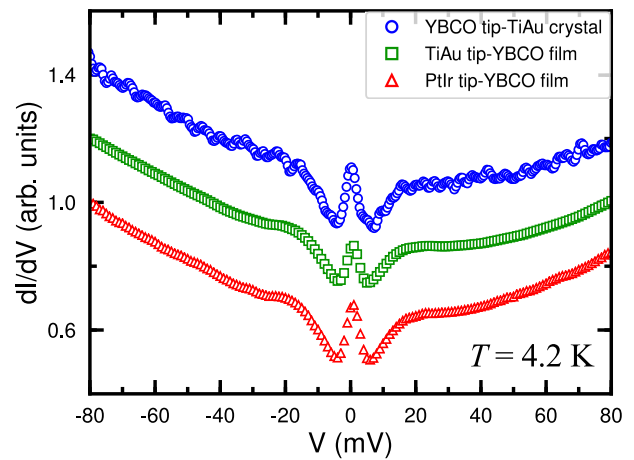}
\caption{\label{fig:TiAu_STS} Tunneling spectra taken between YBCO and TiAu, as well as a reference PtIr tip-(110) YBCO film spectrum, shifted for clarity. The TiAu-YBCO spectra show a ZBC peak of comparable height as the PtIr-YBCO spectrum.
}
\end{figure}

We next show that similar point-contact spectra are obtained with natively-oxidized Nb tips on (110)-oriented YBCO films, as well as TiAu tip-YBCO film junctions. Figure \ref{fig:YBCO_PCS_Nb_TiAu} shows normalized point-contact spectra taken between YBCO and both TiAu and oxidized Nb samples. Both the Nb oxides (squares and triangles) and TiAu (pentagons) exhibit a strong ZBC dip, in contrast to the ZBC peak in the YBCO-Ag spectrum (circles). The stronger ZBC dip for the Nb tip-YBCO film spectrum (triangles) compared to the YBCO tip-Nb foil spectrum (squares) is consistent with the different preparation procedures used for Nb. That is, when used as a film it was sputtered with 3 keV Ar ions, which XPS showed to significantly thin the native Nb$_2$O$_5$ layer (figure \ref{fig:XPS}d). In contrast, when used as a tip the Nb was natively oxidized, so it likely had a relatively thick Nb$_2$O$_5$ layer, similar to the natively-oxidized Nb foil XPS data (figure \ref{fig:XPS}a). As figure \ref{fig:YBCO_PCS_Nb_Ag}(a) showed, Nb samples with thicker Nb$_2$O$_5$ layers exhibit stronger AR suppression, consistent with the above observations.

Having shown that TiAu- and NbO$_x$-YBCO point-contact junctions exhibit AR suppression, we now examine conductance spectra from TiAu-YBCO tunnel junctions, shown in figure \ref{fig:TiAu_STS} along with a reference PtIr-YBCO junction. The PtIr-YBCO spectrum (triangles) exhibits a prominent ZBC peak flanked by side peaks near $\pm$ 20 mV, as expected from $d$-wave Andreev resonance. The TiAu-YBCO data (squares and circles) show a ZBC peak of similar size to the PtIr-YBCO junction. The fact that we observe ZBC peaks for TiAu tunnel junctions indicates that AR suppression is the origin of the ZBC dip in the YBCO-TiAu point-contact junctions.

\begin{figure}
\includegraphics[width=0.45\textwidth]{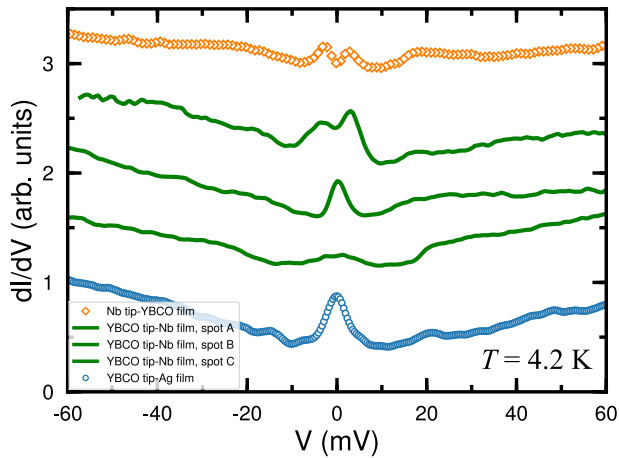}
\caption{\label{fig:Nb_STS} Tunneling spectra taken between YBCO and NbO$_x$, as well as a reference YBCO tip-Ag film spectrum, shifted for clarity. The Nb film was sputtered with 3 keV Ar ions. Compared to the YBCO-Ag film spectra, the YBCO-NbO$_x$ spectra show strong ZBC peaks, weak ZBC peaks, and ZBC peaks split by 6-8 mV.}
\end{figure}

Finally, we examine conductance spectra from YBCO-NbO$_x$ tunnel junctions, shown in figure \ref{fig:Nb_STS} along with a reference YBCO-Ag junction. The YBCO-Ag spectrum (circles) again exhibits a prominent ZBC peak. The YBCO-NbO$_x$ data (lines and diamonds) show a variety of ZBC features depending on the tip's $(x,y)$ position on the sample, including a strong ZBC peak, weak ZBC peak, and ZBC peaks split by 6-8 mV. Again, the fact that we observe ZBC peaks for these NbO$_x$ tunnel junctions provides further evidence that AR suppression is the origin of the ZBC dips in YBCO-NbO$_x$ point-contact junctions.

We now discuss our point-contact AR data, which show that both TiAu and NbO$_x$ suppress $d$-wave AR. To understand this suppression, we consider the interaction of AR and Q-reflection in $d$-wave superconductor/IAFM (dSC/IAFM) junctions, as described in Refs. \cite{Andersen2005,Bobkova2005}. For no interfacial tip/sample barrier, the ZBC peak from $d$-wave Andreev resonance is predicted to be suppressed \cite{Andersen2005}, in which the quasiparticles Q-reflected from the IAFM can re-enter the dSC and disrupt the Andreev resonance. This disruption of AR initially would seem to agree with our observed ZBC dips. However, a finite interfacial barrier is present in all experimental junctions, and such junctions are predicted to show ZBC peaks as well as asymmetric peaks at other energies \cite{Andersen2005}, in disagreement with our data. This disagreement is likely related to discrepancies between the assumptions used in the theoretical work in Ref. \cite{Andersen2005} and our measurements\textemdash namely, the orientation of the IAFM lattice with respect to the sample normal direction, and the relative size of the magnetic and superconducting order parameters in the IAFM and dSC, respectively. Regardless, our observed sensitivity of $d$-wave AR to itinerant antiferromagnetism in point-contact junctions is qualitatively consistent with the Q-reflection mechanism. Theoretical models that allow for arbitrary orientation of the IAFM lattice with respect to the sample normal direction would help in more accurately modeling our point-contact data. Considering that the energies of the aforementioned asymmetric peaks are predicted to depend on the size of the magnetic order parameter in the IAFM, such theoretical models could potentially be utilized to extract quantitative information about the antiferromagnetic ordering in IAFMs from point-contact measurements like ours.

In understanding our YBCO-NbO$_x$ point-contact and tunneling spectra, one factor to consider is the Nb$_{12}$O$_{29}$ content in the NbO$_x$ samples, since the former is the only Nb oxide that is a candidate IAFM. On the one hand, while our NbO$_x$ samples likely have regions of Nb$_{12}$O$_{29}$, they are certainly not pure Nb$_{12}$O$_{29}$, as shown by our XPS data and noting that a specific annealing procedure is required to isolate this compound in bulk form \cite{Cava1991}. However, since Nb$_{12}$O$_{29}$ is the most conductive oxide \footnote{Second to NbO, which is less prevalent in all of our NbO$_x$ samples except the 930 $^{\circ}$C vacuum-annealed foil as shown by the XPS data.} in table \ref{tab:NbOx}, our YBCO-NbO$_x$ junctions are likely preferentially probing Nb$_{12}$O$_{29}$ regions. Furthermore, since AR is only known to be sensitive to itinerant moments, our AR data are likely preferentially sensitive to Nb$_{12}$O$_{29}$. To clarify whether our YBCO-NbO$_x$ data are strongly influenced by the presence of other NbO$_{x}$ compounds, it would be useful to perform point-contact and tunneling measurements between YBCO and pure Nb$_{12}$O$_{29}$ crystals. 

One interesting feature of both our TiAu- and NbO$_x$-YBCO point-contact data is its temperature dependence. Surprisingly, both types of junctions show AR suppression all the way up to YBCO's $T_c$ of 90 K (TiAu also shows suppression up to $T_c$, not shown). If the AR suppression was solely due to Q-reflection, one would reasonably expect the AR suppression to subside near the N\'eel temperature $T_N$ of TiAu (36 K) and Nb$_{12}$O$_{29}$ (12 K). Thus, our observed AR suppression above $T_N$ implies that antiferromagnetic ordering is not necessary to suppress AR. Rather, spin fluctuations may be able to suppress AR. Spin fluctuations are thought to be important in IAFMs \cite{Hasegawa1974}, and indeed, the low $T_N$ and small moment of 0.15 $\mu_B$ in TiAu are consistent with the presence of spin fluctuations \cite{Svanidze2015}. To investigate the sensitivity of $d$-wave AR to spin fluctuations, it might be useful to extend the models from Refs. \cite{Andersen2005,Bobkova2005} to include the effects of disorder in the AFM lattice. For NbO$_x$, if Nb$_{12}$O$_{29}$ is indeed an IAFM, then spin fluctuations would likely also be important. Furthermore, the local moments from oxygen vacancies in Nb$_2$O$_5$ are known to be sources of localized charge and spin \cite{Proslier2008,Proslier2008a,Proslier2009,Proslier2011} and can antiferromagnetically exchange-scatter tunneling electrons, which has been measured up to 40 K and is expected to persist to higher temperatures \cite{Proslier2011,Cao2014}. While it is not experimentally established that AR is affected by local moments, it seems plausible that AR could be upset via spin-flip effects from local moments. More theoretical work is needed to investigate this possibility.

Our tunnel junctions involving TiAu show no AR suppression, which is qualitatively consistent with the Q-reflection work in Refs. \cite{Andersen2002,Andersen2005,Bobkova2005}. Namely, the ZBC peak from $d$-wave Andreev resonance is only predicted to be suppressed in IAFM/dSC junctions with no interfacial barrier, in which the Q-reflected quasiparticles in the IAFM can re-enter the dSC and disrupt the Andreev resonance. For tunnel junctions with large interfacial barriers, the re-entrance of the Q-reflected quasiparticles into the dSC is expected to be negligible, and thus, the IAFM would have little effect on the ZBC peak height, consistent with our data. In contrast to TiAu, tunnel junctions involving NbO$_x$ data show variation in AR suppression, exhibiting ZBC peaks of varying heights as well as ZBC peaks split by 6-8 mV. This spectral variation is consistent with spatial variation of the surface oxide composition. That is, since tunnel junctions are atomic-scale \cite{Granstrom2018} compared to $\sim$ nm-scale point-contact junctions, tunnel junctions can more easily find regions with less Nb$_{2}$O$_{5-x}$ where AR can apparently occur more robustly. However, it is not obvious in the first place why NbO$_x$ would cause weakened or split ZBC peaks for tunnel junctions. The split peaks are not due to the SC gap of Nb $\Delta_{\mathrm{Nb}}$, since $\Delta_{\mathrm{Nb}}\sim$ 1 meV and thermally broadens to only 2.5 mV (e.g. see Ref. \cite{Granstrom2016}). In any case, the peak separation is 6-8 mV for both NbO$_x$ tip-YBCO sample and YBCO tip-NbO$_x$ sample combinations, indicating that it is not sensitive to sample preparation procedure. Since the split and weakened peaks are only observed for NbO$_x$ and not TiAu, it would seem they are related to local oxygen vacancies. To check this observation, it would again be useful to perform tunneling measurements between YBCO and pure Nb$_{12}$O$_{29}$. If the split and weakened ZBC peaks were due to local oxygen vacancies, pure Nb$_{12}$O$_{29}$ would exhibit no such peaks.

In summary, we have used $d$-wave AR spectroscopy with YBCO to probe itinerant antiferromagnetism in  TiAu and NbO$_x$ samples. XPS was also used on the latter to measure their surface oxide composition. For NbO$_x$, samples with a greater degree of oxidation exhibited greater suppression of $d$-wave AR. For both TiAu and NbO$_x$, low-impedance tip-sample junctions suppressed AR more than high-impedance junctions. Furthermore, AR suppression was observed above the N\'eel temperature in both compounds, implying that spin fluctuations can suppress $d$-wave AR. Our data demonstrate that $d$-wave AR is suppressed by itinerant antiferromagnetism, and suggest that $d$-wave AR spectroscopy may be utilized as a nanoscale probe to gauge the degree of spin ordering in IAFMs.

We thank Piotr Bartnicki and Yvette De Sereville for laboratory assistance. We are grateful to Tianhan Liu and Peng Xiong from Florida State University, and Zhijie Chen and Kai Liu from UC Davis for supplying the Nb films.

\bibliography{./Bibliography}

\end{document}